# Chiral symmetry and O(a) improvement in lattice QCD

Martin Lüscher[a], Stefan Sint[b], Rainer Sommer[c] and Peter Weisz[b]

[a] Deutsches Elektronen-Synchrotron DESY,
Notkestrasse 85, D-22603 Hamburg, Germany

[b] Max-Planck-Institut für Physik,
Föhringer Ring 6, D-80805 München, Germany

[c] CERN, Theory Division, CH-1211 Genève 23, Switzerland

## Abstract

The dominant cutoff effects in lattice QCD with Wilson quarks are proportional to the lattice spacing $a$. In particular, the isovector axial current satisfies the PCAC relation only up to such effects. Following a suggestion of Symanzik, they can be cancelled by adding local O($a$) correction terms to the action and the axial current. We here address a number of theoretical issues in connection with the O($a$) improvement of lattice QCD and then show that chiral symmetry can be used to fix the coefficients multiplying the correction terms.

May 1996

## 1. Introduction

A well-known deficit of Wilson's formulation of lattice QCD is that the chiral symmetry of the theory is not preserved [1]. This is usually not considered to be a fundamental problem, because there is every reason to expect that the symmetry is restored in the continuum limit. But one should bear in mind that numerical simulations of lattice QCD are limited to relatively large lattice spacings $a$. Chiral symmetry may then still be rather strongly violated by lattice effects.

A clear demonstration of this has recently been given in ref. [2]. The tests reported there have been performed for the case of quenched QCD with two mass-degenerate light quarks. In the continuum limit of this theory the isovector axial current $A_\mu^a$ is expected to satisfy the PCAC relation

$$\partial_\mu A_\mu^a = 2mP^a, \qquad (1.1)$$

where $P^a$ denotes the associated axial density and $m$ the quark mass. By considering matrix elements of eq. (1.1) between various low-energy states it was found that the lattice corrections to the relation are far from being negligible on the accessible lattices.

With the current simulation algorithms one needs at least a factor 32 more computer time to be able to reduce the lattice spacing by a factor 2 at constant statistical errors and physical length scales. At the same time the lattice effects (which are of order $a$ in the Wilson theory) are only reduced by 50%. In view of this unfavourable situation it is evidently desirable to develop better strategies to bring the cutoff effects under control.

A comparatively simple and theoretically attractive possiblity is to apply Symanzik's improvement programme [3,4], where the O($a$) cutoff effects in on-shell quantities (particle energies, scattering amplitudes, normalized matrix elements of local composite fields between particle states, etc.) are cancelled by adding local O($a$) counterterms to the lattice action and to the composite fields of interest [5–8]. In the so constructed "on-shell improved" theory the continuum limit is reached much faster (with a rate proportional to $a^2$), while the computational cost of the terms added remains small [9–12].

On lattices without boundaries only one counterterm, the Sheikholeslami-Wohlert term [6], needs to be included in the lattice action and not many more are required to improve the low-dimensional operators such as the axial current and density [2,8]. A technical difficulty however is that the coefficients multiplying these counterterms are not a priori known. They are functions of the gauge coupling and should be chosen so that the O($a$) cutoff effects in on-shell quantities cancel.

The main observation made in this paper is that the coefficient $c_{\rm sw}$ multiplying the Sheikholeslami-Wohlert term can be determined by studying the violations of the PCAC relation (1.1) on the lattice. Chiral symmetry restoration and O($a$) improvement thus get tied up in an interesting way. The idea works out both in perturbation theory and non-perturbatively, but here we only set up the theoretical framework and defer all detailed computations to two separate papers [23,24]. A short account of the method has already appeared in ref. [2].

The present paper also gives us the opportunity to discuss the problem of on-shell O($a$) improvement from a somewhat novel point of view (sects. 2 and 3). In particular, the rôle played by the O($a$) counterterms proportional to the quark mass is clarified, and it is shown that mass-independent renormalization schemes must be set up with care if O($a$) improvement is to be preserved.

The matrix elements used to study chiral symmetry restoration are constructed from the QCD Schrödinger functional [13–15,17] which we introduce in sect. 4. For completeness we also derive the form of the O($a$) boundary counterterms that must be included in the action to improve the Schrödinger functional (sect. 5), although this is not really needed when we discuss the lattice corrections to the PCAC relation (1.1) in sect. 6. The paper ends with a few concluding remarks and a series of technical appendices.

## 2. On-shell O($a$) improvement revisited

Our aim in this section is to derive the form of the O($a$) counterterms to the lattice action and the axial current and density that are required for on-shell improvement. Although the improved action has previously been obtained by Sheikholeslami and Wohlert [6], we think it is worthwhile to go through the argumentation again in a slightly different way. The extension of the discussion to the Schrödinger functional will then be rather easy (sect. 5).

*2.1 Preliminaries*

In this section we choose to set up the theory on a four-dimensional hyper-cubic euclidean lattice with spacing $a$ and infinite extent in all directions. Most of our notational conventions are collected in appendix A. The gauge group is taken to be $\mathrm{SU}(N)$ and we assume for simplicity that there are $N_{\mathrm{f}}$ flavours of mass-degenerate light quarks, although it would not be very much more difficult to treat the case of light quarks with different masses.

A gauge field $U$ on the lattice is an assignment of a matrix $U(x,\mu) \in \mathrm{SU}(N)$ to every lattice point $x$ and direction $\mu = 0,1,2,3$. Quark and anti-quark fields, $\psi(x)$ and $\overline{\psi}(x)$, reside on the lattice sites and carry Dirac, colour and flavour indices. The (unimproved) lattice action is of the form

$$S[U,\overline{\psi},\psi] = S_{\mathrm{G}}[U] + S_{\mathrm{F}}[U,\overline{\psi},\psi], \qquad (2.1)$$

where $S_{\mathrm{G}}$ denotes the usual Wilson plaquette action and $S_{\mathrm{F}}$ the Wilson quark action. Explicitly we have

$$S_{\mathrm{G}}[U] = \frac{1}{g_0^2} \sum_p \mathrm{tr}\left\{1 - U(p)\right\} \qquad (2.2)$$

with $g_0$ being the bare gauge coupling and $U(p)$ the parallel transporter around the plaquette $p$. The sum runs over all *oriented* plaquettes $p$ on the lattice.

To define the quark action $S_{\mathrm{F}}$ we first introduce the Wilson-Dirac operator

$$D = \tfrac{1}{2}\left\{\gamma_\mu(\nabla_\mu^* + \nabla_\mu) - a\nabla_\mu^*\nabla_\mu\right\}, \qquad (2.3)$$

which involves the gauge covariant lattice derivatives $\nabla_\mu$ and $\nabla_\mu^*$ defined in appendix A. The action then assumes the standard form,

$$S_{\mathrm{F}}[U,\overline{\psi},\psi] = a^4 \sum_x \overline{\psi}(x)(D + m_0)\psi(x), \qquad (2.4)$$

where $m_0$ denotes the bare quark mass.

## 2.2 Local effective theory

Close to the continuum limit the lattice theory defined above may be described in terms of a local effective theory with action

$$S_{\text{eff}} = S_0 + aS_1 + a^2 S_2 + \ldots \tag{2.5}$$

The leading term, $S_0$, is just the action of the continuum theory, while the other terms are to be interpreted as operator insertions in the continuum theory. In his analysis of the cutoff dependence of lattice field theories, Symanzik [3,4] defines the continuum theory using dimensional regularization, but we could also employ a lattice with spacing very much smaller than $a$ to give a precise meaning to $S_0$ and the operator insertions.

The correction terms in the effective action are of the form

$$S_k = \int \mathrm{d}^4 x \, \mathcal{L}_k(x), \tag{2.6}$$

where the lagrangians $\mathcal{L}_k(x)$ are linear combinations of local composite fields of dimension $4 + k$. The dimension counting here includes the (non-negative) powers of the quark mass $m$ by which some of the fields may be multiplied. From the list of all possible such fields only a small subset occurs in the effective action. First of all, since one integrates over the position $x$, one can use partial integration to reduce the number of terms that must be included. The remaining terms must be invariant under gauge transformations and U(1)×SU($N_\text{f}$) flavour rotations. They should also respect the exact discrete symmetries of the lattice theory. This includes all space-time lattice symmetries and the charge conjugation symmetry (appendix B).

Taking these remarks into account one finds that the order $a$ effective lagrangian, $\mathcal{L}_1(x)$, must be a linear combination of the fields

$$\mathcal{O}_1 = \overline{\psi} \sigma_{\mu\nu} F_{\mu\nu} \psi, \tag{2.7}$$

$$\mathcal{O}_2 = \overline{\psi} D_\mu D_\mu \psi + \overline{\psi} \overleftarrow{D}_\mu \overleftarrow{D}_\mu \psi, \tag{2.8}$$

$$\mathcal{O}_3 = m \operatorname{tr} \{F_{\mu\nu} F_{\mu\nu}\}, \tag{2.9}$$

$$\mathcal{O}_4 = m \{\overline{\psi} \gamma_\mu D_\mu \psi - \overline{\psi} \overleftarrow{D}_\mu \gamma_\mu \psi\}, \tag{2.10}$$

$$\mathcal{O}_5 = m^2 \overline{\psi} \psi, \tag{2.11}$$

where $F_{\mu\nu}$ denotes the gauge field strength and $D_\mu$ the gauge covariant partial derivative (cf. appendix A). In subsects. 2.3 and 2.4 we shall come back to this list of fields and show how it can be reduced to essentially one term.

Cutoff effects originate not only from the lattice action but also from the local composite fields that one is interested in. So let us consider some local gauge invariant field $\phi(x)$ constructed from the quark and gluon fields on the lattice. For simplicity we assume that $\phi(x)$ does not mix with other fields under renormalization. We then expect that the connected renormalized $n$-point correlation functions

$$G_n(x_1,\ldots,x_n) = (Z_\phi)^n \langle \phi(x_1)\ldots\phi(x_n)\rangle_{\text{con}} \qquad (2.12)$$

have a well-defined continuum limit, provided the renormalization factor $Z_\phi$ is chosen appropriately and if all points $x_1,\ldots,x_n$ are kept at non-zero (physical) distances from one another.

In the local effective theory the renormalized lattice field $Z_\phi \phi(x)$ is represented by an effective field

$$\phi_{\text{eff}}(x) = \phi_0(x) + a\phi_1(x) + a^2\phi_2(x) + \ldots \qquad (2.13)$$

The fields $\phi_k(x)$ appearing here are linear combinations of local fields with the appropriate dimension and symmetry properties. To order $a$ the lattice correlation functions are then given by

$$G_n(x_1,\ldots,x_n) = \langle \phi_0(x_1)\ldots\phi_0(x_n)\rangle_{\text{con}}$$

$$- a\int \mathrm{d}^4 y\, \langle \phi_0(x_1)\ldots\phi_0(x_n)\mathcal{L}_1(y)\rangle_{\text{con}}$$

$$+ a\sum_{k=1}^n \langle \phi_0(x_1)\ldots\phi_1(x_k)\ldots\phi_0(x_n)\rangle_{\text{con}} + \mathrm{O}(a^2), \qquad (2.14)$$

where the expectation values on the right hand side are to be taken in the continuum theory with action $S_0$. The second term is the contribution of the order $a$ term in the effective action. Note that the integral over $y$ in general diverges at the points $y = x_k$. A subtraction prescription must hence be supplied. The precise way in which this happens is unimportant, because the arbitrariness that one has amounts to a local operator insertion at these points, i.e. to a redefinition of the field $\phi_1(x)$.

It should also be emphasized that not all the dependence on the lattice spacing comes from the explicit factors of $a$ in eq. (2.14). The other source of $a$-dependence are the fields $\phi_1(x)$ and $\mathcal{L}_1(y)$, which are linear combinations of some basis of fields. While the basis elements are independent of $a$, the coefficients need not be so, although they are expected to vary only slowly with $a$. In perturbation theory the coefficients are calculable polynomials in $\ln(a)$.

*2.3 Cutoff dependence of on-shell amplitudes*

Eventually all on-shell quantities in QCD can be extracted from correlation functions of local composite fields. An important point to note is that for this the correlation functions are only required at non-zero physical distances. We would now like to show that the local effective theory can be simplified considerably if attention is restricted to such correlation functions. As an example we discuss the functions $G_n$ introduced above, but the argumentation is readily carried over to correlation functions involving several kinds of fields.

Let us first consider the second term on the right hand side of eq. (2.14). The effective lagrangian $\mathcal{L}_1(y)$ is a linear combination of the fields (2.7)–(2.11). As long as $y$ keeps away from the points $x_k$, the field equations of the continuum theory may be applied and one concludes that certain linear combinations of these fields do not contribute to the correlation function. This remains true after integration over $y$ up to contact terms that arise at the points $x_1, \ldots, x_k$. Any such contact term amounts to an operator insertion, which, in eq. (2.14), can be compensated by a redefinition of the field $\phi_1$ (the dimensionalities and symmetries of the fields involved do not allow for any other form of the contact terms). Since we do not insist on any particular expression for $\phi_1(x)$, but only require that eq. (2.14) holds, we are hence free to apply the field equations to simplify the effective lagrangian.

The linear combinations of the fields (2.7)–(2.11) that vanish are easily found at tree-level of perturbation theory by applying the classical field equations. There are two such relations which allow us to eliminate $\mathcal{O}_2$ and $\mathcal{O}_4$. At non-zero couplings the coefficients in the vanishing linear combinations change, but since we are only interested in reducing the number of basis fields, it is enough to know that the linear relations allow one to express some of them through the other basis elements. This information is invariant under small changes of the coefficients so that, barring singular events, we may take $\mathcal{L}_1$ to be a linear combination of $\mathcal{O}_1$, $\mathcal{O}_3$ and $\mathcal{O}_5$.

Similar arguments may be used to eliminate some of the terms that con-tribute to the field $\phi_1(x)$. The only place in eq. (2.14) where $\phi_1(x)$ appears is the third term on the right hand side. Since all positions $x_1, \ldots, x_n$ are kept at non-zero distances from each other, no contact terms can arise when the field equations are applied in this correlation function. The number of basis fields from which $\phi_1(x)$ is constructed can thus be reduced as in the case of the effective lagrangian.

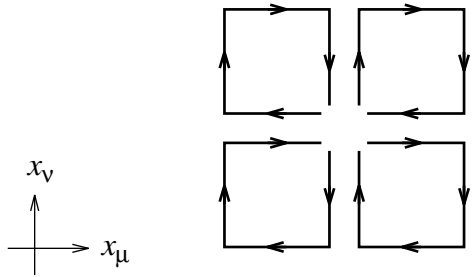

Fig. 1. Graphical representation of the products of gauge field variables contributing to the lattice field strength tensor (2.18). Each square corresponds to one of the terms in eq. (2.19).

*2.4 Improved lattice action*

Our aim is to construct an improved lattice action by adding a suitable $O(a)$ counterterm to the Wilson action (2.1). The counterterm should be chosen such that the order $a$ term in the effective action is cancelled. Since we are only interested in on-shell amplitudes, we may assume that the effective lagrangian $\mathcal{L}_1$ is a linear combination of the fields $\mathcal{O}_1$, $\mathcal{O}_3$ and $\mathcal{O}_5$. It is then quite obvious that $\mathcal{L}_1$ can be made to vanish by adding a counterterm of the form

$$a^5 \sum_x \left\{ c_1 \widehat{\mathcal{O}}_1(x) + c_3 \widehat{\mathcal{O}}_3(x) + c_5 \widehat{\mathcal{O}}_5(x) \right\}, \qquad (2.15)$$

where $\widehat{\mathcal{O}}_k$ is some lattice representation of the field $\mathcal{O}_k$.

Apart from renormalizations of the bare parameters and adjustments of the coefficients $c_k$, the discretization ambiguities that one has here are of order $a^2$. In particular, we may choose to represent the fields $\text{tr}\{F_{\mu\nu}F_{\mu\nu}\}$ and $\overline{\psi}\psi$ by the Wilson plaquette field and the local scalar density that already appears in the Wilson quark action. The $O(a)$ counterterms proportional $\widehat{\mathcal{O}}_3$ and $\widehat{\mathcal{O}}_5$ then amount to a renormalization of the bare coupling and mass. At first sight one might think that such reparametrizations are insignificant, but as explained in sect. 3, this is not quite true if a mass-independent renormalization scheme is employed. For the time being we ignore this complication and simply drop the contributions of $\widehat{\mathcal{O}}_3$ and $\widehat{\mathcal{O}}_5$.

For the on-shell O($a$) improved action we thus obtain

$$S_{\text{impr}}[U,\overline{\psi},\psi] = S[U,\overline{\psi},\psi] + \delta S[U,\overline{\psi},\psi], \tag{2.16}$$

$$\delta S[U,\overline{\psi},\psi] = a^5 \sum_x c_{\text{sw}} \, \overline{\psi}(x) \tfrac{i}{4}\sigma_{\mu\nu} \widehat{F}_{\mu\nu}(x)\psi(x), \tag{2.17}$$

where $S[U,\overline{\psi},\psi]$ is the Wilson action and $\widehat{F}_{\mu\nu}$ a lattice representation of the gluon field strength tensor $F_{\mu\nu}$. We adopt the standard definition

$$\widehat{F}_{\mu\nu}(x) = \frac{1}{8a^2}\left\{Q_{\mu\nu}(x) - Q_{\nu\mu}(x)\right\}, \tag{2.18}$$

$$Q_{\mu\nu}(x) = U(x,\mu)U(x+a\hat{\mu},\nu)U(x+a\hat{\nu},\mu)^{-1}U(x,\nu)^{-1}$$

$$+ U(x,\nu)U(x-a\hat{\mu}+a\hat{\nu},\mu)^{-1}U(x-a\hat{\mu},\nu)^{-1}U(x-a\hat{\mu},\mu)$$

$$+ U(x-a\hat{\mu},\mu)^{-1}U(x-a\hat{\mu}-a\hat{\nu},\nu)^{-1}U(x-a\hat{\mu}-a\hat{\nu},\mu)U(x-a\hat{\nu},\nu)$$

$$+ U(x-a\hat{\nu},\nu)^{-1}U(x-a\hat{\nu},\mu)U(x+a\hat{\mu}-a\hat{\nu},\nu)U(x,\mu)^{-1}. \tag{2.19}$$

The four terms in this equation correspond to the four plaquette loops shown in fig. 1.

The O($a$) counterterm (2.17) has first been written down by Sheikholeslami and Wohlert [6]. In their paper they perform field transformations in the functional integral of the lattice theory to argue that only this term is required for on-shell improvement. Our strategy here has been to achieve a reduction of the number of terms already at the level of the effective action. This treatment is simpler in our opinion, because the discussion of the form of the O($a$) correction terms takes place entirely in the continuum theory.

The coefficient $c_{\text{sw}}$ multiplying the Sheikholeslami-Wohlert term in the improved action is a function of the bare coupling $g_0$ and must be chosen so that the O($a$) cutoff effects in on-shell quantities cancel. It has been shown in ref. [6] that $c_{\text{sw}} = 1$ to lowest order of perturbation theory. The coefficient has later been worked out to one-loop order by Wohlert [7]. His results will be rederived in ref. [23] and a non-perturbative determination of $c_{\rm sw}$, in the quenched approximation and for gauge group $SU(3)$, will be reported in ref. [24].

*2.5 Improved axial current and density*

From the discussion in subsect. 2.2 it is clear that not all $O(a)$ effects can be removed from the correlation functions of local composite fields by employing an improved action. One also has to use improved fields to achieve this. They are constructed in very much the same way as the improved action.

So let us assume that $\phi(x)$ is some given (unimproved) composite field on the lattice. The isovector axial current and density,

$$A_\mu^a(x) = \overline{\psi}(x)\gamma_\mu\gamma_5\tfrac{1}{2}\tau^a\psi(x), \tag{2.20}$$

$$P^a(x) = \overline{\psi}(x)\gamma_5\tfrac{1}{2}\tau^a\psi(x), \tag{2.21}$$

where $\tau^a$ is a Pauli matrix acting on the flavour indices of the quark field, are examples of such fields. We then determine the general form of the order $a$ term, $\phi_1(x)$, in the expansion (2.13) of the associated effective field. Since we are only interested in the improvement of on-shell matrix elements, the classical field equations may be used to eliminate some of the basis fields contributing to $\phi_1(x)$ (cf. subsect. 2.3). The on-shell $O(a)$ improved lattice field is then given by

$$\phi_{\rm I}(x) = \phi(x) + a\delta\phi(x), \tag{2.22}$$

where $\delta\phi(x)$ is obtained by taking the general linear combination of a lattice representation of the remaining basis fields.

Taking into account the transformation behaviour of the axial current (2.20) under the lattice symmetries and charge conjugation, it is straightforward to show that a list of all possible basis fields is in this case given by

$$(\mathcal{O}_6)_\mu^a = \overline{\psi}\gamma_5\tfrac{1}{2}\tau^a\sigma_{\mu\nu}D_\nu\psi - \overline{\psi}\overleftarrow{D}_\nu\sigma_{\mu\nu}\gamma_5\tfrac{1}{2}\tau^a\psi, \tag{2.23}$$

$$(\mathcal{O}_7)_\mu^a = \overline{\psi}\tfrac{1}{2}\tau^a\gamma_5 D_\mu\psi + \overline{\psi}\overleftarrow{D}_\mu\gamma_5\tfrac{1}{2}\tau^a\psi, \tag{2.24}$$

$$(\mathcal{O}_8)_\mu^a = m\overline{\psi}\gamma_\mu\gamma_5\tfrac{1}{2}\tau^a\psi. \tag{2.25}$$

The first of these can be related to the other two by the field equations and so may be dropped. The $O(a)$ counterterm associated with $(\mathcal{O}_8)_\mu^a$ amounts to a renormalization of the axial current. Since we have not imposed any renormalization conditions so far, we may at this point just as well drop this term (the issue will be addressed again in sect. 3).

We are then left with the second term in the list above and conclude that

$$\delta A_\mu^a(x) = c_A \tfrac{1}{2}(\partial_\mu^* + \partial_\mu) P^a(x). \tag{2.26}$$

In the case of the axial density a similar analysis shows that

$$\delta P^a(x) = 0, \tag{2.27}$$

where we have again ignored a mass-dependent renormalization factor.

The coefficient $c_A$ depends on the gauge coupling and is to be chosen so as to achieve the desired improvement. In perturbation theory $c_A$ is of order $g_0^2$, because at tree-level the local axial current (2.20) is already on-shell improved (up to the mass-dependent renormalization factor mentioned above). This has previously been noted in ref. [8]. The coefficient is worked out to one-loop order of perturbation theory in ref. [23] and it is also possible to compute it non-perturbatively, using numerical simulations [24].

## 3. Mass-independent renormalization schemes

In this section we address the problem of the proper parametrization of the improved theory and clarify the rôle played by the order $a$ terms appearing in the local effective theory that correspond to renormalizations of the bare parameters and fields.

### 3.1 Renormalization and O(a) improvement

In the improved theory the renormalization conditions on the gauge coupling, the quark mass and the improved composite fields must be chosen with care. What we would like to achieve is that the correlation functions of the renormalized fields, at fixed non-zero physical distances and fixed renormalized coupling and mass, converge to the continuum limit with a rate proportional to $a^2$. Of course this is only possible if the coefficients $c_{sw}$, $c_A$, etc. have been assigned their proper values (which we assume is the case).

An obvious possibility is to impose all renormalization conditions on a set of renormalized correlation functions defined at the same point $(g_0, am_0)$ in the bare parameter space. The renormalized amplitudes in such a scheme are evidently inert against transformations of the bare parameters and rescalings of the bare fields. Without loss the corresponding $O(a)$ counterterms may hence be dropped (as we did in sect. 2). This also means that the complete list of counterterms has been taken into account and that, therefore, no uncancelled $O(a)$ corrections can arise. In other words, renormalization schemes of this type are automatically compatible with $O(a)$ improvement.

A disadvantage of this procedure however is that the renormalized coupling and fields implicitly depend on the quark mass. Mass-independent schemes, where one imposes the renormalization conditions at zero quark mass, are intrinsically simpler and certainly better suited to discuss the scale evolution of the renormalized parameters [22]. The problem then is that the massive theory must be related to the massless theory, a link which is usually established via the bare parameters. As a result reparametrizations of the bare theory can no longer be ignored if $O(a)$ improvement is to be preserved.

These remarks will become clearer below, where we consider two examples for illustration. We shall then set up the general mass-independent renormalization scheme respecting $O(a)$ improvement. Specific schemes are discussed later in this section.

*3.2 Naive mass-independent schemes*

In the plane of bare parameters a critical line

$$m_0 = m_c(g_0) \tag{3.1}$$

is expected to exist, where the physical quark mass vanishes. Our aim is to parametrize the theory around this line. It is then useful to introduce the subtracted mass

$$m_q = m_0 - m_c. \tag{3.2}$$

As an aside we remark that $m_c(g_0)$ depends on how precisely the physical quark mass is defined. Different definitions lead to values of $m_c(g_0)$ that differ by terms of order $a^2$. The issue will be addressed again in subsect. 6.6 and also in ref. [24]. In this section order $a^2$ corrections are considered negligible and a precise definition of the critical bare mass is then not required.

It is common to assume that the renormalized coupling $g_R$ and the renormalized mass $m_R$ in a mass-independent scheme are related to the bare param-

eters through

$$g_{\rm R}^2 = g_0^2 Z_{\rm g}(g_0^2, a\mu), \tag{3.3}$$

$$m_{\rm R} = m_{\rm q} Z_{\rm m}(g_0^2, a\mu), \tag{3.4}$$

where $\mu$ is a normalization mass and

$$Z(g_0^2, a\mu) = 1 + Z^{(1)}(a\mu) g_0^2 + Z^{(2)}(a\mu) g_0^4 + \ldots \tag{3.5}$$

for $Z = Z_{\rm g}$ and $Z = Z_{\rm m}$. We now show that such schemes always lead to uncancelled O($a$) corrections in some renormalized amplitudes.

At $g_0 = 0$ the quarks decouple and their dynamics is described by the free Wilson quark action. According to eqs. (3.4) and (3.5) (and since $m_{\rm c} = 0$ at zero coupling) the renormalized quark mass $m_{\rm R}$ is equal to the bare mass $m_0$ in this situation. We are thus supposed to take the continuum limit at fixed $m_0$, but as is well-known this leads to uncancelled O($a$) corrections in various places. A particularly obvious case is the "pole mass",

$$m_{\rm p} = \frac{1}{a} \ln(1 + am_0) = m_{\rm R} - \tfrac{1}{2} a m_{\rm R}^2 + \ldots, \tag{3.6}$$

which is equal to the energy of a free quark with zero momentum. It is possible to correct for this deficit by replacing eq. (3.4) through

$$m_{\rm R} = m_{\rm q} \left(1 - \tfrac{1}{2} a m_{\rm q}\right) + {\rm O}(g_0^2). \tag{3.7}$$

The pole mass $m_{\rm p}$ then coincides with $m_{\rm R}$ up to terms of order $a^2$.

One might hope to get away with a modification of eq. (3.4), as suggested above, but it turns out that eq. (3.3) cannot be valid either if O($a$) corrections in renormalized amplitudes are to be avoided. The argumentation is more difficult in this case, because the problem shows up only at one-loop order of perturbation theory. A relatively simple quantity to consider is the running coupling $\bar{g}^2$ introduced in ref. [18]. $\bar{g}^2$ is obtained from the QCD Schrödinger functional by computing the response of the functional to a change of the boundary values of the gauge field. We do not need to know any further details about the coupling here except that it is a well-defined function of the spatial extent $L$ of the lattice, the lattice spacing $a$ and the bare parameters $g_0$ and $m_0$.

The perturbation expansion

$$\bar{g}^2 = g_0^2 + \left[c^{(1,0)} + c^{(1,1)} N_{\mathrm{f}}\right] g_0^4 + \ldots \tag{3.8}$$

has been worked out in ref. [18]. The O($a$) counterterms required for the improvement of the Schrödinger functional have been taken into account in this calculation (cf. sect. 5). If we insert the definition (3.7) of the renormalized quark mass, the result assumes the form

$$c^{(1,0)} = \frac{11N}{24\pi^2} \ln(L/a) + k_1 + \mathrm{O}(a^2), \tag{3.9}$$

$$c^{(1,1)} = -\frac{1}{12\pi^2} \ln(L/a) + k_2 + k_3 a m_{\mathrm{R}} + \mathrm{O}(a^2), \tag{3.10}$$

where $k_2$ is a function of $m_{\mathrm{R}} L$ and

$$k_3 = 0.012000(2). \tag{3.11}$$

We can now express $\bar{g}^2$ as a series in the renormalized coupling $g_{\mathrm{R}}^2$ by eliminating the bare coupling. The renormalization factor $Z_{\mathrm{g}}^{(1)}(a\mu)$ should be chosen so as to cancel the logarithmic divergence, but since it may not depend on the quark mass, we cannot get rid of the term proportional to $k_3$. We thus end up with an uncancelled O($a$) correction and so conclude that eq. (3.3) must be modified to be compatible with O($a$) improvement.

### 3.3 Improved mass-independent schemes

From the discussion in sect. 2 we now recall that a complete O($a$) improvement of the theory in general requires a renormalization of the bare parameters by factors of the form $1 + b(g_0^2) a m_{\mathrm{q}}$. We thus introduce a modified bare coupling and bare quark mass through

$$\tilde{g}_0^2 = g_0^2 \left(1 + b_{\mathrm{g}} a m_{\mathrm{q}}\right), \tag{3.12}$$

$$\widetilde{m}_{\mathrm{q}} = m_{\mathrm{q}} \left(1 + b_{\mathrm{m}} a m_{\mathrm{q}}\right). \tag{3.13}$$

The coefficients $b_{\mathrm{g}}$ and $b_{\mathrm{m}}$ depend on $g_0^2$ and should be chosen so as to cancel any remaining cutoff effects of order $a$. Note that the modified and ordinary bare coupling coincide along the critical line $m_0 = m_{\mathrm{c}}$. The general mass-independent renormalization scheme, compatible with O($a$) improvement, is

now given by

$$g_{\mathrm{R}}^2 = \tilde{g}_0^2 Z_{\mathrm{g}}(\tilde{g}_0^2, a\mu), \qquad (3.14)$$

$$m_{\mathrm{R}} = \widetilde{m}_{\mathrm{q}} Z_{\mathrm{m}}(\tilde{g}_0^2, a\mu), \qquad (3.15)$$

where $Z_{\mathrm{g}}$ and $Z_{\mathrm{m}}$ have an expansion of the form (3.5) with $g_0$ replaced by $\tilde{g}_0$.

It may be worthwhile at this point to discuss the significance of the modified bare parameters $\tilde{g}_0$ and $\widetilde{m}_{\mathrm{q}}$ a little further. Our aim is to parametrize the bare theory in such a way that the continuum limit can be reached coherently for all quark masses and without $\mathrm{O}(a)$ corrections. To approach the limit the bare parameters have to be scaled in a particular way. The basic observation is that the scaling required for $g_0$ necessarily depends on the quark mass (if $\mathrm{O}(a)$ corrections are to be avoided), while $\tilde{g}_0$ scales independently of the quark mass. A similar comment applies to the bare quark mass $m_{\mathrm{q}}$, which must be scaled by a mass-dependent factor. The modified bare mass $\widetilde{m}_{\mathrm{q}}$, on the other hand, is scaled by a factor depending on $a\mu$ only. It should be clear from these remarks that the coefficients $b_{\mathrm{g}}$ and $b_{\mathrm{m}}$ are well-determined and independent of the particular renormalization scheme chosen.

So far we have been exclusively concerned with the parameter renormalization, but it is now straightforward to extend the discussion to the renormalization of multiplicatively renormalizable local fields. Suppose $\phi(x)$ is such a field and let $\phi_{\mathrm{I}}(x)$ be the associated improved field [eq. (2.22)]. Following the conventions adopted in subsect. 2.5, the $\mathrm{O}(a)$ counterterm which amounts to a renormalization of the field by a factor of the form $1 + b(g_0^2) a m_{\mathrm{q}}$ is not included in $\phi_{\mathrm{I}}(x)$. It seems more natural to us to include this factor in the definition

$$\phi_{\mathrm{R}}(x) = Z_\phi(\tilde{g}_0^2, a\mu)(1 + b_\phi a m_{\mathrm{q}})\phi_{\mathrm{I}}(x) \qquad (3.16)$$

of the renormalized field. The coefficient $b_\phi$ plays a rôle completely analogous to $b_{\mathrm{g}}$ and $b_{\mathrm{m}}$. In particular, it is independent of the renormalization condition chosen to fix $Z_\phi$.

In perturbation theory the $b$–coefficients can be expanded according to

$$b = b^{(0)} + b^{(1)} g_0^2 + b^{(2)} g_0^4 + \ldots \qquad (3.17)$$

From the discussion in subsect. 3.2 we infer that $b_{\mathrm{g}}^{(0)} = 0$ and

$$b_{\mathrm{g}}^{(1)} = 0.012000(2) \times N_{\mathrm{f}}. \qquad (3.18)$$

This suggests that the modified bare coupling $\tilde{g}_0^2$ is very nearly equal to $g_0^2$, for all couplings of interest and quark masses $am_q$ less than say 0.1. In practice the difference can probably be ignored until very precise calculations become feasible. It should be noted, incidentally, that $b_g = 0$ in the quenched approximation, because the purely gluonic observables (such as $\bar{g}^2$) are automatically improved at non-zero quark mass if they are at zero quark mass.

In the case of the coefficients $b_m$, $b_A$ and $b_P$ only the tree-level result

$$b_m^{(0)} = -\tfrac{1}{2}, \qquad (3.19)$$

$$b_A^{(0)} = b_P^{(0)} = 1, \qquad (3.20)$$

is currently available [8,23]. The corrections associated with these coefficients may be non-negligible at the larger quark masses.

*3.4 Renormalization conditions*

A complete specification of a mass-independent renormalization scheme requires that we fix the finite parts of the renormalization constants $Z_g$, $Z_m$ and $Z_\phi$ by imposing an appropriate set of renormalization conditions. Different schemes are then related by transformations of the form

$$g_R^2 \to g_R^2 X_g(g_R^2), \qquad m_R \to m_R X_m(g_R^2), \qquad \phi_R \to \phi_R X_\phi(g_R^2) \qquad (3.21)$$

(up to corrections of order $a^2$).

For perturbation theory minimal subtraction is a technically attractive renormalization prescription. It is defined by the requirement that the expansion coefficients $Z_g^{(l)}$, $Z_m^{(l)}$ and $Z_\phi^{(l)}$ are polynomials in $\ln(a\mu)$ with no constant term, to any order $l \geq 1$ of perturbation theory.

At the non-perturbative level mass-independent renormalization schemes are not as easy to define. For such a scheme to be practically useful, the renormalization constants $Z_g$, $Z_m$ and $Z_\phi$ should be computable through numerical simulations. Since they refer to properties of the theory at zero quark mass, one is then confronted with the problem of simulating QCD with very light (or even massless) quarks.

A scheme where this difficulty is bypassed has been described in ref. [2]. The proposition is to impose all renormalization conditions on a set of correlation functions derived from the QCD Schrödinger functional. In this framework an infrared cutoff is provided by the finite extent $L$ of the lattice (given in physical units) and one may then safely set the quark mass to zero without running into singular quark propagators. The philosophy which goes along with this approach has been explained in ref. [2] and a possible choice of renormalization conditions is described there.

## 4. Schrödinger functional

Most of the details given in this section have previously appeared in the literature [13–15,17]. Our aim is not to motivate the use of the Schrödinger functional or to explain its basic properties. For this the reader should consult the papers quoted above and also refs. [20,21], where an alternative approach (using the temporal gauge and spectral boundary conditions for the quark and anti-quark fields) is discussed. Instead we would like to briefly recall the relevant definitions and to set up the notations that will be used in the rest of this paper and in refs. [23–25]. We first introduce the Schrödinger functional without $O(a)$ improvement and derive the required correction terms in sect. 5.

### 4.1 Lattice geometry and fields

In general the formulation of the theory is as in sect. 2 except that the lattice is now taken to be of finite extent in all directions. The possible values of the time coordinate $x_0$ of a lattice point $x$ are then $x_0 = 0, a, 2a, \ldots, T$. At fixed times the lattice is thought to be wrapped on a torus of size $L^3$. In other words, all fields are assumed to be periodic functions of the space coordinates with period $L$.

At the boundaries $x_0 = 0$ and $x_0 = T$ we impose Dirichlet boundary conditions. In the case of the gauge field the requirement is that

$$U(x,k)|_{x_0=0} = W(\mathbf{x},k), \qquad U(x,k)|_{x_0=T} = W'(\mathbf{x},k), \qquad (4.1)$$

where $W(\mathbf{x},k)$ and $W'(\mathbf{x},k)$ are some externally prescribed fields. More precisely, we choose a smooth continuum gauge field $C_k(\mathbf{x})$ and set

$$W(\mathbf{x},k) = \mathcal{P}\exp\left\{ a\int_0^1 dt\, C_k(\mathbf{x}+a\hat{\mathbf{k}}-ta\hat{\mathbf{k}}) \right\}. \qquad (4.2)$$

Similarly, $W'(\mathbf{x},k)$ is given by another field $C'_k(\mathbf{x})$. In eq. (4.2) the symbol $\mathcal{P}$ implies a path-ordered exponential such that the fields at the larger values of $t$ come first. Note that the temporal gauge field variables $U(x,0)$ (which are defined for $0 \leq x_0 < T$) remain unconstrained.

The dynamical degrees of freedom of the quark and anti-quark fields $\psi(x)$ and $\overline{\psi}(x)$ reside on the lattice sites $x$ with $0 < x_0 < T$. At the boundaries only half of the Dirac components are defined and these are fixed to some prescribed values $\rho, \ldots, \bar{\rho}'$. Explicitly, if we introduce the projectors $P_\pm = \frac{1}{2}(1 \pm \gamma_0)$, the boundary conditions on the quark field are

$$P_+\psi(x)|_{x_0=0} = \rho(\mathbf{x}), \qquad P_-\psi(x)|_{x_0=T} = \rho'(\mathbf{x}), \qquad (4.3)$$

while for the anti-quark field we require that

$$\overline{\psi}(x)P_-|_{x_0=0} = \bar{\rho}(\mathbf{x}), \qquad \overline{\psi}(x)P_+|_{x_0=T} = \bar{\rho}'(\mathbf{x}). \qquad (4.4)$$

For consistency the boundary values must be such that the complementary components $P_-\rho, \ldots, \bar{\rho}' P_-$ vanish.

*4.2 Action*

In the interior of the lattice the action density is given by the same expressions as on the infinite lattice considered previously. We however need to specify the precise form of the action in the vicinity of the boundaries $x_0 = 0$ and $x_0 = T$. In particular, for the Wilson plaquette action we now write

$$S_\mathrm{G}[U] = \frac{1}{g_0^2} \sum_p w(p) \operatorname{tr}\{1 - U(p)\}, \qquad (4.5)$$

where the sum runs over all oriented plaquettes $p$ whose corners have time coordinates $x_0$ in the range $0 \leq x_0 \leq T$. The weight $w(p)$ is equal to 1 for all $p$ except for the spatial plaquettes at $x_0 = 0$ and $x_0 = T$, which are given the weight $\frac{1}{2}$.

To be able to write the quark action in an elegant form it is useful to extend the fields to all times $x_0$ by "padding" with zeros. In the case of the quark field this amounts to setting

$$\psi(x) = 0 \quad \text{if } x_0 < 0 \text{ or } x_0 > T, \qquad (4.6)$$

and

$$P_-\psi(x)|_{x_0=0} = P_+\psi(x)|_{x_0=T} = 0. \qquad (4.7)$$

The anti-quark field is extended similarly and the so far undefined link variables are set to 1. With this notational convention the Wilson-Dirac operator (2.3) is well-defined at all times and the quark action is again given by eq. (2.4).

An important detail we should mention at this point is that an unconventional phase factor $\lambda_\mu$ has been included in the definition (A.13)–(A.17) of the covariant lattice derivatives $\nabla_\mu$ and $\nabla_\mu^*$. This factor is equivalent to imposing the generalized periodic boundary conditions

$$\psi(x + L\hat{k}) = \mathrm{e}^{i\theta_k}\psi(x), \qquad \overline{\psi}(x + L\hat{k}) = \overline{\psi}(x)\mathrm{e}^{-i\theta_k}, \tag{4.8}$$

as one easily proves by performing an abelian gauge transformation. The present formulation, where the fields are strictly periodic and $\lambda_\mu$ appears in the difference operators, is technically simpler. In any case, the angles $\theta_k$ parametrize a family of admissible boundary conditions and give us further opportunities to probe the quark dynamics.

*4.3 Functional integral and correlation functions*

The Schrödinger functional

$$\mathcal{Z}[C', \overline{\rho}', \rho'; C, \overline{\rho}, \rho] = \int \mathrm{D}[U]\mathrm{D}[\psi]\mathrm{D}[\overline{\psi}]\,\mathrm{e}^{-S[U,\overline{\psi},\psi]} \tag{4.9}$$

involves an integration over all fields with the specified boundary values. Note that the integration measure does not depend on the boundary values of the fields. All dependence on the latter derives from the action.

The expectation value of any product $\mathcal{O}$ of fields is given by

$$\langle \mathcal{O} \rangle = \left\{ \frac{1}{\mathcal{Z}} \int \mathrm{D}[U]\mathrm{D}[\psi]\mathrm{D}[\overline{\psi}]\,\mathcal{O}\,\mathrm{e}^{-S[U,\overline{\psi},\psi]} \right\}_{\overline{\rho}'=\rho'=\overline{\rho}=\rho=0}. \tag{4.10}$$

Apart from the gauge field and the quark and anti-quark fields integrated over, $\mathcal{O}$ may involve the "boundary fields"

$$\zeta(\mathbf{x}) = \frac{\delta}{\delta\overline{\rho}(\mathbf{x})}, \qquad \overline{\zeta}(\mathbf{x}) = -\frac{\delta}{\delta\rho(\mathbf{x})}, \tag{4.11}$$

$$\zeta'(\mathbf{x}) = \frac{\delta}{\delta\overline{\rho}'(\mathbf{x})}, \qquad \overline{\zeta}'(\mathbf{x}) = -\frac{\delta}{\delta\rho'(\mathbf{x})}. \tag{4.12}$$

These are perfectly meaningful in eq. (4.10) (the derivatives act on the Boltzmann factor) and have the effect of inserting certain combinations of $\psi(x)$ and $\overline{\psi}(x)$ close to the boundaries of the lattice, together with the appropriate gauge field variables to ensure gauge covariance (see appendix C for the precise definition of the variational derivatives).

## 5. O($a$) improvement of the Schrödinger functional

The reader may wish to skip this section on first reading, because the discussion of the PCAC relation in sect. 6 is only marginally dependent on the results obtained here. For a more solid understanding of our approach it is however useful to know that the Schrödinger functional is well-behaved in the continuum limit and that the improvement programme extends to this quantity in a straightforward manner. A further motivation to include this section is that we would like to prepare the ground for the computation of the running coupling and the running quark mass along the lines described in ref. [2].

In the case of the pure gauge theory, the problem of the renormalization and O($a$) improvement of the Schrödinger functional has already been addressed in subsects. 2.5 and 4.5 of ref. [15]. The argumentation presented there carries over literally to QCD with any number of quarks. We shall, therefore, restrict the discussion to those aspects that we believe are new or worthwhile to be reconsidered.

### 5.1 Continuum limit and scope of improvement

In the following the boundary values $C$, $C'$, $\rho, \ldots, \overline{\rho}'$ are taken to be linear combinations of a finite number of plane waves with coefficients and momenta that are kept fixed as the lattice spacing is sent to zero. Since we assume periodic boundary conditions in the space directions, the momenta must be integer multiples of $2\pi/L$.

As for the correlation functions we restrict attention to the expectation values of products of local composite fields $\phi(x)$ and the Fourier components of the boundary quark fields $\zeta(\mathbf{y}), \ldots, \overline{\zeta}'(\mathbf{z})$. We assume that the fields $\phi(x)$ are inserted at non-zero (physical) distances from the boundaries and from each other.

The renormalization of the Schrödinger functional is achieved by expressing the bare parameters through the renormalized parameters and by scaling the boundary values of the quark and anti-quark field with a renormalization constant $Z_\zeta$ [15,17]. For fixed renormalized boundary values, as specified above, the continuum limit can then be taken and one ends up with a well-defined functional of the boundary values of the gauge and quark fields. By differentiating with respect to the latter, the continuum limit of the correlation functions of the Fourier components of the boundary fields $\zeta(\mathbf{y}), \ldots, \bar{\zeta}'(\mathbf{z})$ is obtained at the same time. Operator insertions in the interior of the space-time volume require additional renormalizations as in infinite volume.

It should be noted, however, that the operator insertions and the boundary fields are treated quite differently. In the first case the insertions are made at positions that are at non-zero distances from each other and from the boundaries. A discussion of short distance singularities is thus avoided, and the renormalizations needed are just those that are already required in on-shell matrix elements of the operators. The correlation functions of the boundary fields, on the other hand, are obtained in momentum space by differentiation of the Schrödinger functional. Arbitrary (finite) products of the Fourier components of these fields can occur. Short distance singularities associated with such products are automatically taken care of on the level of the Schrödinger functional, where they would show up in the form of divergent local boundary terms, constructed from the boundary values of the fields.

Our aim in this section is to study the cutoff dependence of the correlation functions of the type mentioned above and to determine the required $\mathrm{O}(a)$ counterterms. The improvement of the fields $\phi(x)$ is as in infinite volume and will not be considered again. We are then left with the correlation functions of the boundary quark fields. As in the case of the renormalization of the theory, it is simpler to discuss the improvement of the Schrödinger functional itself. Once the latter is improved the correlation functions will be improved, too. In particular, no further subtractions will be needed, and the $\mathrm{O}(a)$ counterterms $\delta\zeta(\mathbf{y}), \ldots, \delta\bar{\zeta}'(\mathbf{z})$ are all equal to zero.

### 5.2 Boundary terms and effective theory

The approach of the Schrödinger functional to the continuum limit can be described by a local effective theory. Compared to our discussion in sect. 2 of the infinite volume theory the principal difference is that we now need to include further terms in the effective action to account for boundary effects. The general form of the correction terms in eq. (2.5) then is

$$S_k = \int \mathrm{d}^4 x\, \mathcal{L}_k(x) + \lim_{\epsilon \to 0} \int \mathrm{d}^3 x\, \left\{ \mathcal{B}_k(x)|_{x_0=\epsilon} + \mathcal{B}'_k(x)|_{x_0=T-\epsilon} \right\}, \qquad (5.1)$$

where $\mathcal{B}_k(x)$ and $\mathcal{B}'_k(x)$ are linear combinations of local composite fields of dimension $3 + k$. A technical remark we should make here is that the limit $\epsilon \to 0$ is non-trivial in general and requires that the coefficients in the linear combinations are scaled in a particular way [13,14]. This complication is not very important in the present context, because we are only interested in the general form of the boundary terms, which is determined by symmetry considerations. The relevant symmetries are all internal symmetries of the lattice theory and the discrete space rotations and reflections. Moreover $\mathcal{B}_k(x)$ and $\mathcal{B}'_k(x)$ are related by a time reflection so that only one of them needs to be discussed.

In subsect. 2.3 we have shown that under certain conditions the field equations may be used to reduce the number of terms contributing to the effective action. The same arguments may be applied here, first to simplify the effective lagrangian $\mathcal{L}_1(x)$ and then also the boundary terms $\mathcal{B}_1(x)$ and $\mathcal{B}'_1(x)$. The limit $\epsilon \to 0$ is helpful at this point, because the applicability of the field equations to reduce the number of terms is made obvious. As for the volume term, $\mathcal{L}_1(x)$, we conclude that it can be taken to be the same as in the infinite volume theory. For the improved lattice action we thus write

$$S_{\text{impr}}[U, \overline{\psi}, \psi] = S[U, \overline{\psi}, \psi] + \delta S_{\text{v}}[U, \overline{\psi}, \psi] + \delta S_{\text{G,b}}[U] + \delta S_{\text{F,b}}[U, \overline{\psi}, \psi], \quad (5.2)$$

$$\delta S_{\text{v}}[U, \overline{\psi}, \psi] = a^5 \sum_{x_0=a}^{T-a} \sum_{\mathbf{x}} c_{\text{sw}} \overline{\psi}(x) \tfrac{i}{4} \sigma_{\mu\nu} \widehat{F}_{\mu\nu}(x) \psi(x), \quad (5.3)$$

where the notation is as in subsect. 2.4 (the index "v" indicates a volume counterterm, while the boundary counterterms are labeled by "b"). In the following we deduce the form of the boundary counterterms.

Table 1. Improvement coefficients $c_{\text{s,t}}^{(k,l)}$

| $N$ | $c_{\text{s}}^{(1,0)}$ | $c_{\text{t}}^{(1,0)}$ | $c_{\text{t}}^{(1,1)}$ | $c_{\text{t}}^{(2,0)}$ |
|---|---|---|---|---|
| 2 | $-0.166(1)$ [15] | $-0.0543(5)$ [15] | $0.0191410(1)$[18] | $-0.0115(5)$ [19] |
| 3 |  | $-0.08900(5)$ [16] | $0.0191410(1)$[18] |  |

*5.3 Boundary counterterms independent of the quark fields*

In the pure gauge theory any gauge invariant local composite field has dimension greater or equal to 4. The only fields that can contribute to $\mathcal{B}_1$ and $\mathcal{B}_1'$ are hence given by

$$\mathcal{O}_9 = \text{tr}\,\{F_{kl}F_{kl}\}, \tag{5.4}$$

$$\mathcal{O}_{10} = \text{tr}\,\{F_{0k}F_{0k}\}. \tag{5.5}$$

For the associated O($a$) counterterm we may take [15]

$$\delta S_{\text{G,b}}[U] = \frac{1}{2g_0^2}(c_{\text{s}} - 1)\sum_{p_{\text{s}}}\text{tr}\,\{1 - U(p_{\text{s}})\}$$

$$+ \frac{1}{g_0^2}(c_{\text{t}} - 1)\sum_{p_{\text{t}}}\text{tr}\,\{1 - U(p_{\text{t}})\}. \tag{5.6}$$

The sums here run over all oriented plaquettes $p_{\text{s}}$ and $p_{\text{t}}$ at the boundaries that are space-like ($p_{\text{s}}$) or time-like ($p_{\text{t}}$). The notation is otherwise as in sect. 2.

In perturbation theory we have

$$c_{\text{s,t}} = 1 + c_{\text{s,t}}^{(1)}g_0^2 + c_{\text{s,t}}^{(2)}g_0^4 + \ldots, \tag{5.7}$$

$$c_{\text{s,t}}^{(k)} = \sum_{l=0}^{k} c_{\text{s,t}}^{(k,l)}N_{\text{f}}^l. \tag{5.8}$$

The known expansion coefficients for $N = 2$ and $N = 3$ are listed in table 1 (the numbers in square brackets point to the references from which the coefficients have been taken).

### 5.4 Boundary counterterms depending on the quark fields

The presence of the quark fields allows one to construct many more composite fields of dimension 4 with the required symmetries. A basis of such fields is given by

$$\mathcal{O}_{11} = \overline{\psi} P_+ D_0 \psi + \overline{\psi} \overleftarrow{D}_0 P_- \psi, \tag{5.9}$$

$$\mathcal{O}_{12} = \overline{\psi} P_- D_0 \psi + \overline{\psi} \overleftarrow{D}_0 P_+ \psi, \tag{5.10}$$

$$\mathcal{O}_{13} = \overline{\psi} P_+ \gamma_k D_k \psi - \overline{\psi} \overleftarrow{D}_k \gamma_k P_- \psi, \tag{5.11}$$

$$\mathcal{O}_{14} = \overline{\psi} P_- \gamma_k D_k \psi - \overline{\psi} \overleftarrow{D}_k \gamma_k P_+ \psi, \tag{5.12}$$

$$\mathcal{O}_{15} = m \overline{\psi} \psi. \tag{5.13}$$

The formal field equations imply two relations,

$$\mathcal{O}_{11} + \mathcal{O}_{13} + \mathcal{O}_{15} = 0, \tag{5.14}$$

$$\mathcal{O}_{12} - \mathcal{O}_{14} - \mathcal{O}_{15} = 0, \tag{5.15}$$

so that two fields can be eliminated. A particularly simple form of the associated boundary counterterms to the lattice action is obtained if we choose

$$\mathcal{O}_{11}, \mathcal{O}_{14}, \mathcal{O}_{15} \quad \text{at} \quad x_0 = 0, \tag{5.16}$$

$$\mathcal{O}_{12}, \mathcal{O}_{13}, \mathcal{O}_{15} \quad \text{at} \quad x_0 = T, \tag{5.17}$$

as the basis of fields. Note that we are free to choose a different basis at the two boundaries of the lattice.

The counterterm corresponding to $\mathcal{O}_{15}$ is of a special kind, similar to the volume counterterms associated with $\mathcal{O}_3$ and $\mathcal{O}_5$ (cf. sect. 2). To see this we consider the term at $x_0 = 0$ and choose the lattice field, representing $\mathcal{O}_{15}$, to be given by

$$m_{\text{q}} \left\{ \overline{\psi}(y) U(x,0)^{-1} P_+ \psi(x) + \overline{\psi}(x) P_- U(x,0) \psi(y) \right\}_{y=x+a\hat{0}}. \tag{5.18}$$

Next we note that the terms in the Wilson quark action that are linear in the boundary values $\rho$ and $\bar{\rho}$ are precisely proportional to the above expression.

The boundary counterterm can thus be compensated by rescaling $\rho$ and $\bar\rho$ by a factor of the form $1 + b(g_0^2)am_{\mathrm{q}}$. There is also a term in the action depending quadratically on the boundary values, but since it is already of $\mathrm{O}(a)$ the rescaling of this term produces a correction of $\mathrm{O}(a^2)$ which is considered negligible in the present discussion.

Since one anyway has to renormalize the boundary values of the quark and anti-quark fields, the $\mathrm{O}(a)$ counterterms corresponding to $\mathcal{O}_{15}$ are hence not included in the improved action. If a mass-independent renormalization scheme is employed, the counterterm instead appears in the definition

$$\zeta_{\mathrm{R}}(\mathbf{x}) = Z_\zeta(\tilde g_0^2, a\mu)(1 + b_\zeta a m_{\mathrm{q}})\zeta(\mathbf{x}) \tag{5.19}$$

of the renormalized boundary field $\zeta_{\mathrm{R}}(\mathbf{x})$ (the other boundary fields are renormalized similarly).

We are thus left with altogether four boundary counterterms, two at time $x_0 = 0$ and two at $x_0 = T$. Their coefficients must be such that the time reversal invariance of the theory is preserved. A possible choice of the counterterms then is

$$\delta S_{\mathrm{F,b}}[U,\overline\psi,\psi] = a^4 \sum_{\mathbf{x}} \Big\{ (\tilde c_{\mathrm{s}} - 1)\big[\widehat{\mathcal{O}}_{\mathrm{s}}(\mathbf{x}) + \widehat{\mathcal{O}}'_{\mathrm{s}}(\mathbf{x})\big]$$
$$+ (\tilde c_{\mathrm{t}} - 1)\big[\widehat{\mathcal{O}}_{\mathrm{t}}(\mathbf{x}) - \widehat{\mathcal{O}}'_{\mathrm{t}}(\mathbf{x})\big] \Big\}, \tag{5.20}$$

where

$$\widehat{\mathcal{O}}_{\mathrm{s}}(\mathbf{x}) = \tfrac{1}{2}\bar\rho(\mathbf{x})\gamma_k(\nabla_k^* + \nabla_k)\rho(\mathbf{x}), \tag{5.21}$$

$$\widehat{\mathcal{O}}'_{\mathrm{s}}(\mathbf{x}) = \tfrac{1}{2}\bar\rho'(\mathbf{x})\gamma_k(\nabla_k^* + \nabla_k)\rho'(\mathbf{x}), \tag{5.22}$$

$$\widehat{\mathcal{O}}_{\mathrm{t}}(\mathbf{x}) = \big\{\overline\psi(y)P_+\nabla_0^*\psi(y) + \overline\psi(y)\overleftarrow\nabla_0^* P_-\psi(y)\big\}_{y=(a,\mathbf{x})}, \tag{5.23}$$

$$\widehat{\mathcal{O}}'_{\mathrm{t}}(\mathbf{x}) = \big\{\overline\psi(y)P_-\nabla_0\psi(y) + \overline\psi(y)\overleftarrow\nabla_0 P_+\psi(y)\big\}_{y=(T-a,\mathbf{x})}. \tag{5.24}$$

Note that one does not refer to any "undefined" components of the quark and anti-quark fields in these expressions. The projectors $P_\pm$ and the time derivatives are always such that this is avoided.

As will be shown in ref. [23] the Schrödinger functional in the free quark theory with Wilson action is already O($a$) improved. The perturbation expansion of the improvement coefficients $\tilde{c}_\text{s}$ and $\tilde{c}_\text{t}$ is hence of the form

$$\tilde{c}_{\text{s},\text{t}} = 1 + \tilde{c}^{(1)}_{\text{s},\text{t}} g_0^2 + \tilde{c}^{(2)}_{\text{s},\text{t}} g_0^4 + \ldots \qquad (5.25)$$

It may seem a bit artificial to write the coefficients in eq. (5.20) in the way we did. The notation was chosen to emphasize the analogy with the pure gauge counterterm (5.6). Moreover the terms appearing in eq. (5.20) combine with the terms in the Wilson quark action. At $x_0 = 0$, for example, we have

$$a\widehat{\mathcal{O}}_\text{t}(\mathbf{x}) = \overline{\psi}(x+a\hat{0})\psi(x+a\hat{0})$$
$$- \overline{\psi}(x+a\hat{0})U(x,0)^{-1}P_+\psi(x) - \overline{\psi}(x)P_-U(x,0)\psi(x+a\hat{0}). \quad (5.26)$$

This counterterm thus contributes to the mass term at $x_0 = a$ and changes the weight of the time-like hopping terms at the boundary from 1 to $\tilde{c}_\text{t}$.

# 6. Chiral symmetry restoration

We now proceed to study the axial current conservation on the lattice. The coefficients $c_{\text{sw}}$, $c_{\text{A}}$, etc. are initially assumed to have their proper values so that the theory is on-shell $\text{O}(a)$ improved. Chiral symmetry in lattice QCD with Wilson quarks has been extensively discussed in the literature (an incomplete list of references is [26–29]). Here we show that $\text{O}(a)$ improvement leads to a significant reduction of the cutoff effects in the PCAC relation. This may conversely be taken as an "improvement condition" to fix $c_{\text{sw}}$ and $c_{\text{A}}$.

*6.1 PCAC relation*

In a mass-independent renormalization scheme, the renormalized improved axial current and density are given by

$$(A_{\text{R}})_\mu^a(x) = Z_{\text{A}}(\tilde{g}_0^2, a\mu)(1 + b_{\text{A}} a m_{\text{q}})\left\{A_\mu^a(x) + \delta A_\mu^a(x)\right\}, \tag{6.1}$$

$$(P_{\text{R}})^a(x) = Z_{\text{P}}(\tilde{g}_0^2, a\mu)(1 + b_{\text{P}} a m_{\text{q}})P^a(x) \tag{6.2}$$

(cf. subsect. 2.5). The renormalization constants $Z_{\text{A}}$ and $Z_{\text{P}}$ should be fixed by imposing appropriate renormalization conditions. In what follows we do not need to know exactly how this is done. We shall however take it for granted that the normalization mass $\mu$ is independent of the lattice size (this is the normal situation, but the remark is made here, because in the SF scheme discussed in ref. [2] one sets $\mu = 1/L$).

We now first consider the theory in infinite volume. Since the PCAC relation (1.1) holds in the continuum limit, and since the improvement has been fully implemented, we expect that

$$\left\langle \tfrac{1}{2}(\partial_\mu^* + \partial_\mu)(A_{\text{R}})_\mu^a(x)\,\mathcal{O}\right\rangle = 2m_{\text{R}}\left\langle (P_{\text{R}})^a(x)\,\mathcal{O}\right\rangle + \text{O}(a^2), \tag{6.3}$$

for any product $\mathcal{O}$ of renormalized improved fields located at non-zero distances from $x$ and from each other. As already indicated by the notation, the proportionality constant $m_{\text{R}}$ is a renormalized quark mass which can be shown to be of the form (3.15) up to corrections of order $a^2$.

There is one aspect of eq. (6.3) which may appear to be trivial but which will be most important in what follows. The property we wish to emphasize is that the relation holds for any product $\mathcal{O}$ of fields localized in a region not containing $x$. Of course this is just a reflection of the fact that the PCAC relation becomes an operator identity when the theory is formulated in Minkowski space. Different choices of $\mathcal{O}$ then correspond to considering different matrix elements of the relation.

It may nevertheless be useful to briefly recall the standard proof of eq. (6.3) in the euclidean framework. First note that the lattice correlation functions on both sides of the equation converge to the continuum limit with a rate proportional to $a^2$. It is hence sufficient to establish eq. (6.3) in the continuum theory. To this end one performs a local infinitesimal chiral transformation on the quark fields integrated over in the functional integral. We can choose the transformation to be trivial outside a small neighbourhood $R$ of the point $x$. The variation $\delta S_0$ of the action is proportional to

$$\int_R \mathrm{d}^4 y \; \omega^a(y) \left\{ \partial_\mu (A_\mathrm{R})^a_\mu(y) - 2 m_\mathrm{R} (P_\mathrm{R})^a(y) \right\}, \tag{6.4}$$

where $\omega^a(y)$ is the (position dependent) transformation parameter. The product $\mathcal{O}$ is localized outside the region $R$ and is hence not affected by the transformation. The invariance of the quark integration measure then implies

$$\langle \delta S_0 \, \mathcal{O} \rangle = 0, \tag{6.5}$$

which reduces to eq. (6.3) after passing to the local limit.

Exactly the same argumentation may be applied if the theory is set up in a finite volume with Schrödinger functional boundary conditions. We are thus led to conclude that eq. (6.3) must be valid in this situation too, with the same value of $m_\mathrm{R}$, provided the point $x$ is at a non-zero distance from the space-time boundaries. At first sight it may not be totally obvious that the proportionality constant $m_\mathrm{R}$ is indeed independent of the lattice sizes $T$ and $L$ (up to corrections of order $a^2$). A second derivation of this important fact is therefore given in appendix D.

We conclude this subsection by noting that eq. (6.3) remains valid if $\mathcal{O}$ is taken to be a product of unimproved fields or if the O($a$) boundary counterterms $\delta S_\mathrm{G,b}$ and $\delta S_\mathrm{F,b}$ are dropped. In other words, to ensure that the lattice corrections in eq. (6.3) are of order $a^2$, it is enough to include the Sheikholeslami-Wohlert term in the action and to insert the renormalized improved axial current and density.

The reason for this is best explained by considering a simple example. Let us assume that $\mathcal{O}$ is equal to some renormalized improved field $\phi_\mathrm{R}(y)$ and suppose we drop the O($a$) counterterm $a\delta\phi(y)$. From the discussion of the local effective theory in subsect. 2.2 we then infer that eq. (6.3) receives an order $a$ correction given by

$$a\left\langle\left\{\partial_\mu(A_{\rm R})^a_\mu(x) - 2m_{\rm R}(P_{\rm R})^a(x)\right\}\phi_1(y)\right\rangle. \tag{6.6}$$

The correlation function in this formula is to be evaluated in the continuum theory. Since the PCAC relation holds exactly in this limit, the correlation function vanishes and we conclude that there is in fact no order $a$ correction. It is important here that $y$ keeps away from $x$ as otherwise a non-zero contact term can arise which would invalidate the argument.

*6.2 Choice of boundary values*

By varying the boundary values of the fields and by considering various field products $\mathcal{O}$, the PCAC relation can be probed in very many different ways. We now make some definite choices to reduce this enormous flexibility to just a few parameters. In doing so we are guided by the requirement of simplicity and by practical considerations.

As in refs. [15,16] the boundary values $C$ and $C'$ of the gauge field are taken to be constant diagonal matrices. Explicitly, we set

$$C_k = \frac{i}{L}\begin{pmatrix} \phi_1 & 0 & \cdots & 0 \\ 0 & \phi_2 & \cdots & 0 \\ \vdots & \vdots & \ddots & \vdots \\ 0 & 0 & \cdots & \phi_N \end{pmatrix}, \tag{6.7}$$

where the angles $\phi_\alpha$ are required to satisfy

$$\sum_\alpha \phi_\alpha = 0, \qquad \phi_1 < \phi_2 < \ldots < \phi_N, \qquad \phi_N - \phi_1 < 2\pi. \tag{6.8}$$

The boundary value $C'$ at $x_0 = T$ is similarly given by another set of angles $\phi'_\alpha$. For such boundary values one can prove [15] that the gauge field action is minimized by the constant colour-electric field with field strength

$$F_{0k} = (C'_k - C_k)/T, \qquad F_{kl} = 0. \tag{6.9}$$

In many respects this field plays the rôle of a classical background field.

A good feature of the chosen boundary values $C$ and $C'$ is that the translation invariance in the space directions is preserved. The Fourier components of the boundary quark fields may hence be interpreted as operators that create quarks and anti-quarks with definite momenta. In particular, the product

$$\mathcal{O} = a^6 \sum_{\mathbf{y},\mathbf{z}} \bar{\zeta}_{\mathrm{R}}(\mathbf{y}) \gamma_5 \tfrac{1}{2} \tau^a \zeta_{\mathrm{R}}(\mathbf{z}) \qquad (6.10)$$

creates a quark and an anti-quark with zero momenta at time $0$. The correlation functions in eq. (6.3) then are proportional to the probability amplitude that the quark anti-quark pair propagates to the interior of the space-time volume and that it annihilates at the point $x$ (see fig. 2). As usual such pictures have a precise meaning if the quark lines are thought to represent quark propagators at the current gauge field. The derivation of the corresponding Feynman rules from the functional integral is straightforward and will be discussed in some detail in ref. [23].

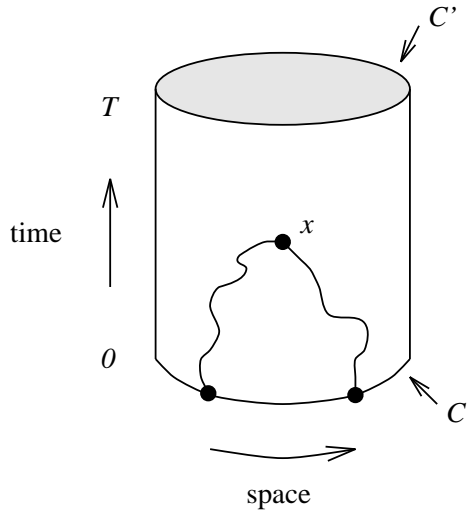

Fig. 2. Two-dimensional sketch of the space-time manifold. $C$ and $C'$ are the boundary values of the gauge field at time $0$ and $T$, respectively. The irregular lines represent the space-time trajectories of a quark and an anti-quark, which are created at time $0$ by the operator (6.10) and annihilate each other at the point $x$.

With the choices made so far the kinematical parameters left to play with are the lattice sizes $T$ and $L$, the time $x_0$ at which the axial current is inserted, the angles $\phi_\alpha$ and $\phi'_\alpha$ characterizing the boundary values of the gauge field and the angles $\theta_k$ appearing in the quark action (cf. appendix A). The latter could be traded for a shift of the spatial quark momenta by the vector $(\theta_1, \theta_2, \theta_3)/L$, but as already explained in sect. 4 we prefer to formulate the theory in terms of strictly periodic fields and to consider the angles $\theta_k$ as some parameters that allow us to probe the quark dynamics in an interesting way.

*6.3 Tests of chiral symmetry*

The size of the error term in eq. (6.3) (with $C$, $C'$ and $\mathcal{O}$ as specified above) can be taken as a measure for the violation of chiral symmetry on the lattice. In the following lines we show how to separate the lattice effects from the universal terms in the equation. A strategy to calculate the coefficients $c_{\text{sw}}$ and $c_{\text{A}}$ will then emerge.

We begin by introducing the bare correlation functions

$$f_{\text{A}}(x_0) = -a^6 \sum_{\mathbf{y},\mathbf{z}} \tfrac{1}{3} \langle A_0^a(x)\, \bar\zeta(\mathbf{y})\gamma_5 \tfrac{1}{2}\tau^a \zeta(\mathbf{z}) \rangle, \qquad (6.11)$$

$$f_{\text{P}}(x_0) = -a^6 \sum_{\mathbf{y},\mathbf{z}} \tfrac{1}{3} \langle P^a(x)\, \bar\zeta(\mathbf{y})\gamma_5 \tfrac{1}{2}\tau^a \zeta(\mathbf{z}) \rangle. \qquad (6.12)$$

A summation over the flavour index is implied here and the O($a$) boundary counterterms are omitted from the improved action (cf. subsect. 6.1). $f_{\text{A}}$ and $f_{\text{P}}$ depend on the bare parameters $g_0$ and $m_0$, the coefficient $c_{\text{sw}}$ at the chosen coupling, and the kinematical parameters listed in subsect. 6.2. There is no dependence on the spatial coordinates of $x$ because of translation invariance.

The PCAC relation (6.3) implies that the time derivative of $f_{\text{A}}$ is proportional to $f_{\text{P}}$ up to cutoff effects. So if we define the ratio

$$m = \tfrac{1}{2}\left[\tfrac{1}{2}(\partial_0^* + \partial_0)f_{\text{A}}(x_0) + c_{\text{A}} a \partial_0^* \partial_0 f_{\text{P}}(x_0)\right]/f_{\text{P}}(x_0), \qquad (6.13)$$

and if we again assume that the coefficients $c_{\text{sw}}$ etc. have their proper values, it is straightforward to show that

$$m_{\text{R}} = m\, \frac{Z_{\text{A}}(1 + b_{\text{A}} a m_{\text{q}})}{Z_{\text{P}}(1 + b_{\text{P}} a m_{\text{q}})} + \text{O}(a^2) \qquad (6.14)$$

[cf. eqs. (6.1),(6.2)]. The ratio $m$ may be regarded as an unrenormalized current quark mass, but it should be noted that $m$ depends on the same parameters as $f_{\text{A}}$ and $f_{\text{P}}$ and additionally on $c_{\text{A}}$.

We now consider two configurations of the kinematical parameters at the same point $(g_0, a m_0)$ in the bare parameter space. The simplest possibility is to insert the axial current at two different times, but we might also choose $C = C' = 0$ in one case and some non-zero $C$ and $C'$ in the other. Let $m_1$ and $m_2$ be the associated values of the unrenormalized current quark mass $m$ defined above. Since $m_R$ and the renormalization factor in eq. (6.14) do not depend on the kinematical parameters, it follows that

$$m_1 - m_2 = \mathrm{O}(a^2). \tag{6.15}$$

By calculating $m_1$ and $m_2$ we thus obtain a direct check on the size of the lattice effects. Note that if one is interested in computing the renormalized quark mass $m_R$ via eq. (6.14), a non-zero value of $m_1 - m_2$ implies a corresponding variation in $m_R$. In other words, $m_R$ is only computable up to such systematic uncertainties. The cutoff effects acquire a concrete meaning in this way which allows us to judge whether they are of any practical importance on the lattices considered.

*6.4 Chiral symmetry violation at tree-level of perturbation theory*

A first impression on the size of the cutoff effects can already be gained at tree-level of perturbation theory. The required computations are straightforward and will be discussed in more detail in ref. [23]. In the limit $g_0 \to 0$ the gauge field is frozen to the minimal action configuration with field strength as given by eq. (6.9). One then has to calculate the quark propagator in this classical background field. The correlation functions $f_A$ and $f_P$ are obtained by taking the product of two propagators with apropriately contracted indices. As already mentioned before, we have $c_{\mathrm{sw}} = 1$ and $c_A = 0$ to lowest order of perturbation theory.

If we choose zero boundary values, $C = C' = 0$, the background field vanishes and the quarks propagate freely. The unrenormalized current quark mass $m$ can be calculated analytically in this case. It turns out that $m$ is independent of the time $x_0$ and that

$$m = m_{\mathrm{p}} + \mathrm{O}(a^2), \tag{6.16}$$

as expected for free quarks [$m_{\mathrm{p}}$ denotes the "pole mass", eq. (3.6)]. The size of the error term in this equation depends on $T$, $L$, $m_0$ and the angles $\theta_k$, but is generally found to be small for reasonable choices of these parameters. On a $16 \times 8^3$ lattice with $am_0 = 0.1$ and $\theta_k = 0$, for example, we have $a(m - m_{\mathrm{p}}) = 0.0006$. Uncertainties in the renormalized quark mass of this order of magnitude are usually considered negligible.

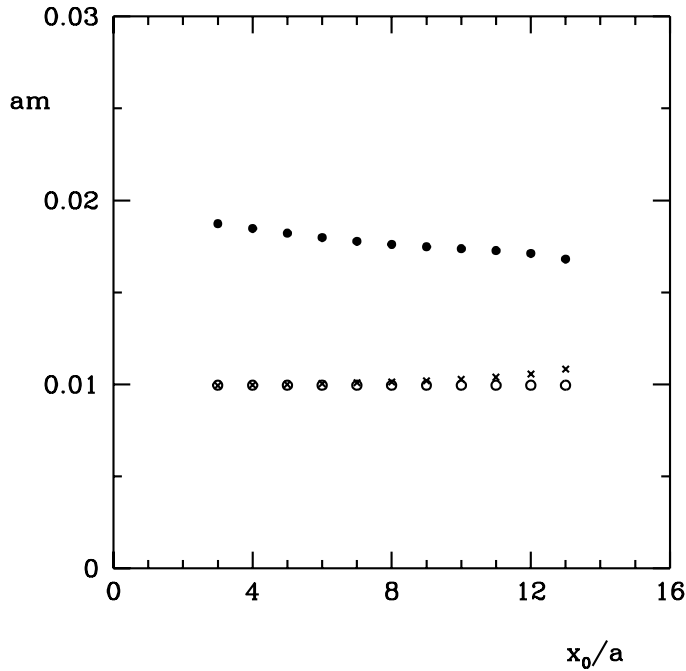

Fig. 3. Plot of $am$ at tree-level of perturbation theory. All data are for a $16 \times 8^3$ lattice with $am_0 = 0.01$ and $\theta_k = 0$. The boundary values of the gauge field are zero (open circles) or as given by eq. (6.17). Full circles and crosses correspond to $c_{\mathrm{sw}} = 0$ and $c_{\mathrm{sw}} = 1$ respectively.

The situation becomes more interesting in the presence of a non-zero background field. To illustrate this we consider the theory with gauge group SU(3) and choose

$$\begin{aligned}(\phi_1, \phi_2, \phi_3) &= \tfrac{1}{6}\left(-\pi, 0, \pi\right), \\ (\phi'_1, \phi'_2, \phi'_3) &= \tfrac{1}{6}\left(-5\pi, 2\pi, 3\pi\right).\end{aligned} \qquad (6.17)$$

As shown in fig. 3 the corresponding values of $m$ now depend on $c_{\mathrm{sw}}$ and also on $x_0$. Without improvement (i.e. for $c_{\mathrm{sw}} = 0$) a very large deviation from the free quark value is observed. Such uncertainties in the quark mass determination are clearly not tolerable. One might suspect that an unreasonably large external scale has been induced into the system by our choice of boundary values for the gauge field. But this is not the case, since the background field has zero frequency and its strength is small in lattice units (on a $16\times 8^3$ lattice we have $|a^2 F^a_{\mu\nu}| \leq 0.028$).

The problem disappears almost completely if we set $c_{\rm sw} = 1$. Improvement thus works very well. The residual cutoff effects seen in fig. 3 are of order $a^2$. They increase significantly towards the boundary $x_0 = T$, but this is the expected behaviour when the distance $T - x_0$, measured in numbers of lattice spacings, becomes small. In the middle of the lattice the differences between the calculated quark masses are again negligible.

### 6.5 Strategy to compute $c_{\rm sw}$ and $c_{\rm A}$

It should be quite obvious now that the coefficients $c_{\rm sw}$ and $c_{\rm A}$ can be calculated by requiring the unrenormalized quark mass $m$ to be independent of the kinematical parameters (up to terms of order $a^2$). We need to consider three different kinematical configurations, corresponding to mass values $m_1$, $m_2$ and $m_3$. The two coefficients are then to be adjusted so that $m_1 = m_2 + {\rm O}(a^2)$ and $m_2 = m_3 + {\rm O}(a^2)$.

A difficulty which becomes apparent at this point is that $c_{\rm sw}$ and $c_{\rm A}$ are not uniquely determined by these equations unless the order $a^2$ corrections are negligible. In perturbation theory the problem is not felt, because the lattice spacing can be made arbitrarily small compared to the external length scales. The coefficients thus have a unique expansion in powers of $g_0$ and the computational strategy described above applies straightforwardly [23].

At non-zero values of $g_0$ one also has the dynamically generated mass scales such as the pion decay constant $F_\pi$. In units of these scales the lattice spacing is a function of $g_0$ so that at fixed coupling there is no way to reduce an error term proportional to say $(aF_\pi)^2$. The coefficients $c_{\rm sw}$ and $c_{\rm A}$ are hence ambiguous by terms of order $aF_\pi$. There is nothing fundamentally wrong with this. The ambiguity simply reflects the fact that O($a$) improvement is an asymptotic concept, valid up to higher-order corrections.

In practice our aim is to cancel the largest contributions to the lattice effects in the accessible range of lattice spacings. This can be achieved by requiring $m_1 = m_2 = m_3$ to hold exactly for a definite set of well-chosen kinematical configurations. Of course one should check that different such "improvement conditions" yield consistent results within small variations. Eventually one *defines* the improved theory by adopting a particular set of improvement conditions. Any residual cutoff effects then still have to be extrapolated away by calculating the quantities of interest on a sequence of lattices with decreasing lattice spacings. A non-perturbative calculation of $c_{\rm sw}$ and $c_{\rm A}$ along these lines will be reported in ref. [24].

*6.6 Cutoff effects and the critical bare mass*

Although O($a$) improvement is very efficient in reducing the lattice effects, one should not forget that chiral symmetry remains to be an only approximate symmetry of the lattice theory. In particular, the critical line $m_c(g_0)$ is not unambiguously defined. This has previously been mentioned and is here discussed in some more detail (cf. subsect. 3.2).

The critical bare mass is loosely defined as the value of $m_0$ at which the physical quark mass vanishes. From chiral perturbation theory one expects that the pion becomes massless at this point and the critical line is hence often characterized in this way. Another option is to define an unrenormalized current quark mass $m$ through a ratio of correlation functions, as in subsect. 6.3, and to search for the value of $m_0$ where $m$ vanishes [2,29]. This method is not restricted to a particular physical situation, because the PCAC relation (6.3) holds for any choice of boundary conditions, field product $\mathcal{O}$ and lattice sizes $T$ and $L$.

The problem now is that the pion mass and the quark masses that one extracts from the PCAC relation do not pass through zero at exactly the same value of $m_0$. As a result of the lattice corrections in eq. (6.3), one rather finds that the calculated critical bare masses vary by terms of order $a^2$ (without improvement the uncertainty would be of order $a$).

In perturbation theory these ambiguities are unimportant, because the lattice spacing can be taken to zero at any finite order of the expansion. At the non-perturbative level some residual cutoff effects are always present, as in the case of the coefficients $c_{\text{sw}}$ and $c_\text{A}$ discussed above. One should also take into account that numerical simulations can only be performed for limited ranges of the external scales. Calculations of the critical bare mass may then also be biased by lattice effects associated with these scales.

## 7. Concluding remarks

So far we have been mainly interested in the case of small quark masses and so did not pay too much attention to the coefficients multiplying the terms proportional to $am_q$ (such as $b_A$ and $b_P$). There are a number of these coefficients and much work will be needed to get them under control. While it is clear how to proceed in perturbation theory, it may not be so easy to disentangle the cutoff effects proportional to $am_q$ from physical quark mass effects at the non-perturbative level, using numerical simulations.

Our discussion of the improvement of the axial current and density in subsect. 2.5 can be readily extended to other local fields. For low-dimensional fields, such as the isovector vector current [2], the $O(a)$ counterterms have a simple form. One may then be able to determine the associated coefficients by requiring the improved fields to transform in the expected way under chiral rotations up to corrections of order $a^2$.

Once the improvement has been fully implemented, the chiral Ward identities may be used to relate different renormalized fields. An interesting case to consider is the scalar density $\overline{\psi}\psi$. This field mixes with the constant field under renormalization. At zero quark mass the subtraction constant and also the multiplicative renormalization constant may now be fixed by insisting that the renormalized scalar density $S_R$ should be related to the axial density $(P_R)^a$ through an infinitesimal (isovector) chiral transformation. The expectation value $\langle S_R \rangle$ then becomes an order parameter for spontaneous chiral symmetry breaking in large volumes.

We would like to thank Hartmut Wittig and Ulli Wolff for a critical reading of a first draft of this paper.

## Appendix A

*A.1 Index conventions*

Lorentz indices $\mu, \nu, \ldots$ are taken from the middle of the Greek alphabet and run from 0 to 3. Latin indices $k, l, \ldots$ run from 1 to 3 and are used to label the components of spatial vectors. For the Dirac indices capital letters $A, B, \ldots$ from the beginning of the alphabet are taken. They run from 1 to 4. Colour vectors in the fundamental representation of SU($N$) carry indices $\alpha, \beta, \ldots$ ranging from 1 to $N$, while for vectors in the adjoint representation, Latin indices $a, b, \ldots$ running from 1 to $N^2 - 1$ are employed. By abuse of notation such indices are also used for the flavour label of the axial current and density.

Repeated indices are always summed over unless otherwise stated and scalar products are taken with euclidean metric.

*A.2 Dirac matrices*

We choose a chiral representation for the Dirac matrices, where

$$\gamma_\mu = \begin{pmatrix} 0 & e_\mu \\ e_\mu^\dagger & 0 \end{pmatrix}. \tag{A.1}$$

The $2 \times 2$ matrices $e_\mu$ are taken to be

$$e_0 = -1, \qquad e_k = -i\sigma_k, \tag{A.2}$$

with $\sigma_k$ the Pauli matrices. It is then easy to check that

$$\gamma_\mu^\dagger = \gamma_\mu, \qquad \{\gamma_\mu, \gamma_\nu\} = 2\delta_{\mu\nu}. \tag{A.3}$$

Furthermore, if we define $\gamma_5 = \gamma_0 \gamma_1 \gamma_2 \gamma_3$, we have

$$\gamma_5 = \begin{pmatrix} 1 & 0 \\ 0 & -1 \end{pmatrix}. \tag{A.4}$$

In particular, $\gamma_5 = \gamma_5^\dagger$ and $\gamma_5^2 = 1$. The hermitean matrices

$$\sigma_{\mu\nu} = \frac{i}{2} [\gamma_\mu, \gamma_\nu] \tag{A.5}$$

are explicitly given by

$$\sigma_{0k} = \begin{pmatrix} \sigma_k & 0 \\ 0 & -\sigma_k \end{pmatrix}, \qquad \sigma_{ij} = -\epsilon_{ijk} \begin{pmatrix} \sigma_k & 0 \\ 0 & \sigma_k \end{pmatrix}, \tag{A.6}$$

where $\epsilon_{ijk}$ is the totally anti-symmetric tensor with $\epsilon_{123} = 1$.

*A.3 Gauge group*

The Lie algebra su$(N)$ of SU$(N)$ can be identified with the space of complex $N \times N$ matrices $X_{\alpha\beta}$ which satisfy

$$X^\dagger = -X, \qquad \text{tr}\,\{X\} = 0, \tag{A.7}$$

where $X^\dagger$ denotes the adjoint matrix of $X$ and $\text{tr}\,\{X\} = X_{\alpha\alpha}$ is the trace of $X$. We may choose a basis $T^a, a = 1, 2, \ldots, N^2 - 1$, in this space such that

$$\text{tr}\,\{T^a T^b\} = -\tfrac{1}{2}\delta^{ab}. \tag{A.8}$$

For $N = 2$, for example, the standard basis is

$$T^a = \frac{\tau^a}{2i}, \quad a = 1, 2, 3, \tag{A.9}$$

where $\tau^a$ denote the Pauli matrices. With these conventions the structure constants $f^{abc}$, defined through

$$[T^a, T^b] = f^{abc} T^c, \tag{A.10}$$

are real and totally anti-symmetric under permutations of the indices.

*A.4 Lattice derivatives*

Ordinary forward and backward lattice derivatives act on colour singlet functions $f(x)$ and are defined through

$$\partial_\mu f(x) = \frac{1}{a}\bigl[f(x + a\hat\mu) - f(x)\bigr], \tag{A.11}$$

$$\partial^*_\mu f(x) = \frac{1}{a}\bigl[f(x) - f(x - a\hat\mu)\bigr], \tag{A.12}$$

where $\hat\mu$ denotes the unit vector in direction $\mu$. The gauge covariant derivative operators, acting on a quark field $\psi(x)$, are given by

$$\nabla_\mu \psi(x) = \frac{1}{a}\bigl[\lambda_\mu U(x, \mu)\psi(x + a\hat\mu) - \psi(x)\bigr], \tag{A.13}$$

$$\nabla^*_\mu \psi(x) = \frac{1}{a}\bigl[\psi(x) - \lambda_\mu^{-1} U(x - a\hat\mu, \mu)^{-1}\psi(x - a\hat\mu)\bigr]. \tag{A.14}$$

The origin of the phase factors

$$\lambda_\mu = e^{ia\theta_\mu/L}, \qquad \theta_0 = 0, \quad -\pi < \theta_k \leq \pi, \qquad (A.15)$$

is explained in subsect. 4.2. They depend on the spatial extent $L$ of the lattice and are all equal to 1 on the infinite lattice. The left action of the lattice derivative operators is defined by

$$\overline{\psi}(x)\overleftarrow{\nabla}_\mu = \frac{1}{a}\left[\,\overline{\psi}(x+a\hat{\mu})U(x,\mu)^{-1}\lambda_\mu^{-1} - \overline{\psi}(x)\,\right], \qquad (A.16)$$

$$\overline{\psi}(x)\overleftarrow{\nabla}_\mu^* = \frac{1}{a}\left[\,\overline{\psi}(x) - \overline{\psi}(x-a\hat{\mu})U(x-a\hat{\mu},\mu)\lambda_\mu\,\right]. \qquad (A.17)$$

*A.5 Continuum gauge fields*

An SU($N$) gauge potential in the continuum theory is a vector field $A_\mu(x)$ with values in the Lie algebra su($N$). It may thus be written as

$$A_\mu(x) = A_\mu^a(x)T^a \qquad (A.18)$$

with real components $A_\mu^a(x)$. The associated field tensor,

$$F_{\mu\nu}(x) = \partial_\mu A_\nu(x) - \partial_\nu A_\mu(x) + [A_\mu(x), A_\nu(x)], \qquad (A.19)$$

may be decomposed similarly and the right and left action of the covariant derivative $D_\mu$ is defined by

$$D_\mu \psi(x) = (\partial_\mu + A_\mu + i\theta_\mu/L)\psi(x), \qquad (A.20)$$

$$\overline{\psi}(x)\overleftarrow{D}_\mu = \overline{\psi}(x)(\overleftarrow{\partial}_\mu - A_\mu - i\theta_\mu/L). \qquad (A.21)$$

The abelian gauge field $i\theta_\mu/L$ appearing here corresponds to the phase factors $\lambda_\mu$ in the lattice theory [eqs. (A.13)–(A.17)].

**Appendix B**

Under charge conjugation the gauge field transforms according to

$$U(x,\mu) \to U(x,\mu)^*. \tag{B.1}$$

The transformation law for the quark and anti-quark fields reads

$$\psi(x) \to \mathcal{C}^{-1}\overline{\psi}(x)^T, \qquad \overline{\psi}(x) \to -\psi(x)^T\mathcal{C}, \tag{B.2}$$

where $\mathcal{C}$ is a $4 \times 4$ matrix satisfying

$$\gamma_\mu^* = -\mathcal{C}\gamma_\mu\mathcal{C}^{-1}. \tag{B.3}$$

If the Dirac matrices are chosen as specified in appendix A, we may take $\mathcal{C} = i\gamma_0\gamma_2$ so that $\mathcal{C}^{-1} = \mathcal{C}^\dagger = \mathcal{C}$.

It follows from these definitions that the Wilson action is invariant under charge conjugation. This is true both on the infinite lattice and for Schrödinger functional boundary conditions. In the latter case the transformation is applied to the field variables at all sites of the lattice including the boundaries $x_0 = 0$ and $x_0 = T$.

**Appendix C**

Let $F[\rho, \bar\rho]$ be a monomial in the boundary values $\rho(\mathbf{x})$ and $\bar\rho(\mathbf{x})$ of degree $d_F$. The Dirac components of the boundary values are not all independent since

$$P_-\rho(\mathbf{x}) = 0 \quad \text{and} \quad \bar\rho(\mathbf{x})P_+ = 0. \tag{C.1}$$

The basic property of the variational derivatives,

$$\left.\frac{\mathrm{d}}{\mathrm{d}t}F[\rho + t\nu, \bar\rho]\right|_{t=0} = a^3 \sum_{\mathbf{x}} (-1)^{d_F - 1} \frac{\delta F}{\delta\rho(\mathbf{x})}\,\nu(\mathbf{x}), \tag{C.2}$$

$$\left.\frac{\mathrm{d}}{\mathrm{d}t}F[\rho, \bar\rho + t\bar\nu]\right|_{t=0} = a^3 \sum_{\mathbf{x}} \bar\nu(\mathbf{x})\frac{\delta F}{\delta\bar\rho(\mathbf{x})}, \tag{C.3}$$

is hence not sufficient to define them uniquely. Uniqueness is achieved by imposing the constraints

$$\frac{\delta F}{\delta \rho(\mathbf{x})} P_- = 0 \quad \text{and} \quad P_+ \frac{\delta F}{\delta \bar\rho(\mathbf{x})} = 0. \tag{C.4}$$

A sum over the Dirac, colour and flavour indices is implicit in eqs. (C.2),(C.3). As usual the definition of the variational derivatives extends to arbitrary polynomials $F[\rho, \bar\rho]$ by linearity. In the same way one also defines the derivatives with respect to the other boundary values $\rho'(\mathbf{x})$ and $\bar\rho'(\mathbf{x})$.

### Appendix D

In this appendix we show that the mass $m_R$ appearing in the PCAC relation (6.3) must be independent of $T$ and $L$ up to corrections of order $a^2$. Since the error term in eq. (6.3) is also of this order, it suffices to consider the continuum theory and to prove that $m_R$ does not depend on $T$ and $L$ in this limit.

We first fix $L$ and discuss what happens when $T$ is changed. Let $\mathcal{P}$ be any product of fields localized in the interior of the space-time manifold. The key observation is that

$$\frac{\partial}{\partial T} \langle \mathcal{P} \rangle_{\mathrm{con}} = - \int \mathrm{d}^3 z \, \langle \mathcal{H}(z) \mathcal{P} \rangle_{\mathrm{con}}, \tag{D.1}$$

where $\mathcal{H}(z)$ denotes the energy density and $z_0$ must be greater than the time coordinates occurring in the product $\mathcal{P}$ and less than $T$. It is straightforward to deduce this from the familiar quantum mechanical representation of the euclidean correlation functions [15,17].

We now consider the expression

$$\mathcal{Q} = \left\{ \partial_\mu (A_R)^a_\mu(x) - 2 m_R (P_R)^a(x) \right\} (P_R)^a(y), \tag{D.2}$$

where $x$ and $y$ are such that $0 < x_0 < y_0 < T$. From the above and eq. (6.3) we then deduce that

$$0 = \frac{\partial}{\partial T} \langle \mathcal{Q} \rangle_{\mathrm{con}}$$

$$= -2 \frac{\partial m_R}{\partial T} \langle (P_R)^a(x)(P_R)^a(y) \rangle_{\mathrm{con}} - \int \mathrm{d}^3 z \, \langle \mathcal{H}(z) \mathcal{Q} \rangle_{\mathrm{con}}. \tag{D.3}$$

The last term in this equation vanishes because of the PCAC relation (6.3) with $\mathcal{O}$ replaced by $\mathcal{H}(z)(P_{\mathrm{R}})^a(y)$. Since the two-point function of the axial density does not vanish for general $x$ and $y$, we conclude that $\partial m_{\mathrm{R}}/\partial T = 0$.

For this argumentation to work out we did not need to refer to any special property of the Schrödinger functional boundary conditions. With periodic boundary conditions in the time direction the same conclusion would be reached. Moreover, since the boundary conditions do not matter in the limit $T \to \infty$, one infers that $m_{\mathrm{R}}$ has to be the same in both cases.

To discuss the dependence of $m_{\mathrm{R}}$ on $L$ we are hence free to choose periodic boundary conditions in all directions. We may then interchange the time with one of the space axes and repeat the argumentation given above to deduce that $m_{\mathrm{R}}$ must be independent of the extent of space-time in this direction, too.